\documentclass[epj, twocolumn]{webofc}
\usepackage[varg]{txfonts}   
\usepackage{multirow}
\woctitle{NSRT2015}
\begin{document}
\title{Microscopic calculations of the characteristics of radiative nuclear reactions for double-magic nuclei}
%
%

\author{Oleg Achakovskiy\inst{1} \fnsep\thanks{\email{oachakovskiy@ippe.ru}} \and
        Sergei Kamerdzhiev\inst{2} \and
        Victor Tselyaev\inst{3}   \and
        Mikhail Shitov\inst{4}
}

\institute{Institute for Physics and Power Engineering,  Obninsk,  Russia
\and
            National Research Centre "Kurchatov Institute", Moscow,  Russia
           \and
           Physical Faculty,  St. Petersburg State University,   St. Petersburg,  Russia
           \and
            National Research Nuclear University MEPhI, Moscow, Russia
          }

\abstract{%
The neutron capture cross sections and   average radiative widths $\Gamma_\gamma$ of neutron resonances for two double-magic nuclei  $^{132}$Sn and $^{208}$Pb have been calculated  using the microscopic photon strength functions,  which were  obtained  within the microscopic self-consistent version of the extended theory of finite Fermi systems  in the  time blocking approximation. For the first time, the microscopic PSFs have been obtained within the fully self-consistent approach  with exact accounting for the single particle continuum (for $^{208}$Pb). The approach includes phonon coupling  effects  in addition to the standard RPA approach. The known  Skyrme force has  been  used. The calculations of nuclear reaction characteristics  have been performed with the EMPIRE 3.1 nuclear reaction code. Here,
 three nuclear level density (NLD) models have been used: the so-called phenomenological  GSM, the EMPIRE specific  (or Enhanced GSM) and the  microscopical  combinatorial HFB NLD models.
For both  considered characteristics we found  a significant disagreement between the results obtained with the GSM and HFB  NLD models. 
 For $^{208}$Pb,  a reasonable  agreement has been found with systematics  for the $\Gamma_\gamma$ values with HFB NLD and with  the experimental data  for  the HFB NLD  average resonance spacing $D_0$,  while  for   these two  quantities  the  differences between the values obtained with GSM and HFB NLD are of several orders of  magnitude. The discrepancies between the results with the phenomenological EGLO PSF  and microscopic RPA or TBA are much less for the same NLD model.   
 }   
\maketitle
\section{Introduction}
\label{intro}

In order to calculate characteristics of nuclear reactions with gamma-rays, 
the information is necessary,  at least,  about the photon strength function (PSF) and nuclear level density  (NLD) models. Traditionally,  these quantities have  been parametrized phenomenologically with the parameters fitted 
for stable nuclei. For example,  the PSF has been parametrized on the basis of smooth Lorentzian type functions but,   as it was noted in \cite{RIPL2,  ripl3},  these  phenomenological Lorentzian-type expressions for PSF are not able to predict the observed structures (under that condition that the Brink-Axel hypothesis is true).  
Also,  the shortcomings of analytical NLD formulae in matching experimental data are overcome,  as a rule,  by empirical parameter adjustments. For these reasons,  the application of  phenomenological models for PSF and NLD to nuclei far from the stability valley is questionable \cite{RIPL2,  ripl3}.

However,  there are also questions of this type  for double-magic nuclei. The problem is that  the phenomenological approaches ''smooth'' the individual characteristics 
of these nuclei or consider them on the average. Individual  peculiarities are especially expressive just for double-magic nuclei,  even for stable,  not to mention unstable those,  whose properties  can be unknown.  For example,  to include the vibrational NLD enhancement    to the well-known so-called generalized superfluid model (GSM) \cite{RIPL2,  ripl3},
the experimental values for the energies of the first 2$^+$  and the formula $50A^{-2/3}$ MeV for the first 3$^-$ levels  are used. The formula is not suited for double-magic nuclei,  and both of these prescriptions  should  not be suited for unstable nuclei.  
  The microscopic approach in the nuclear theory accounts for specificity of each nucleus through its single-particle and collective (phonon) spectra.
Therefore,  it allows  ''some irregular changes'' obtained in the global phenomenological models for nuclear reactions data \cite{Belan} to be seen and checked.  Thus,   for double-magic nuclei   it is necessary  to use the microscopic approaches for both  PSF and NLD. 

In this work, 
 we have applied for PSF the self-consistent version of  microscopic extended theory of finite Fermi systems (ETFFS) \cite{PhysRep}  in the time blocking approximation (TBA) \cite{Tsel2007}. For NLD,  we used the phenomenological GSM \cite{RIPL2},  the EMPIRE specific NLD model \cite{Empire} and the microscopic HFB plus  combinatorial NLD model \cite{Gor2008}. The calculations of  neutron capture cross sections  and average radiative widths of neutron resonances
for two double-magic nuclei -- the stable $^{208}$Pb and unstable $^{132}$Sn -- have been made using the EMPIRE 3.1 nuclear reaction code.   
The comparison with the phenomenological PSF  EGLO model
has been also performed. In all the PSF calculations the smoothing parameter 200 keV has been used. See the details of the calculations  in \cite{ave2011} and \cite{thisconf}.
  
 Quite  recently, the  fully self-consistent calculations of giant resonances  \cite{TselLut} have been realized for double-magic nuclei within both RPA and TBA.
 As a new feature in these calculations,  the single-particle continuum was included, thus avoiding the artificial discretization usually implied in RPA and TBA. In our previous calculations for semi-magic nuclei,  see,  for example,  \cite{ave2011, Acha2015, thisconf}, a discretization procedure for the single particle continuum
 was used (which gave the same results within CRPA for double-magic nuclei \cite{ave2011}).   As  the self-consistent TBA calculations of photoabsorption or PSF are rather time-consuming, 
 we have used the corresponding results for $^{132}$Sn from Ref. \cite{ave2011} and for $^{208}$Pb from \cite{TselLut} obtained with the Skyrme forces SLy4  to  calculate  the above-mentioned nuclear reaction characteristics.


\section{Photon strength functions}
\label{sec-1}

\begin{figure}[t]
\centering
\sidecaption
\includegraphics[width=8cm, clip]{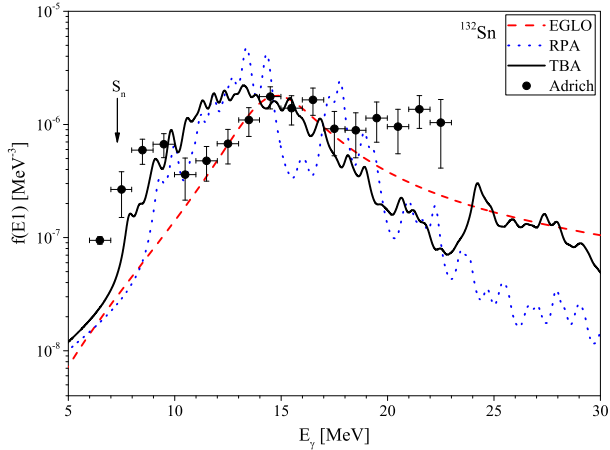}
\caption{Color online. The PSF for $^{132}$Sn. Dotted lines correspond to the self-consistent RPA,  solid lines to the TBA  (including PC),  and dashed lines to the EGLO model \cite{RIPL2}. Experimental data  \cite{Adrich} were recalculated by us for PSF.}
\label{fig-1}       
\end{figure}

\begin{figure}[t]
\centering
\sidecaption
\includegraphics[width=8cm, clip]{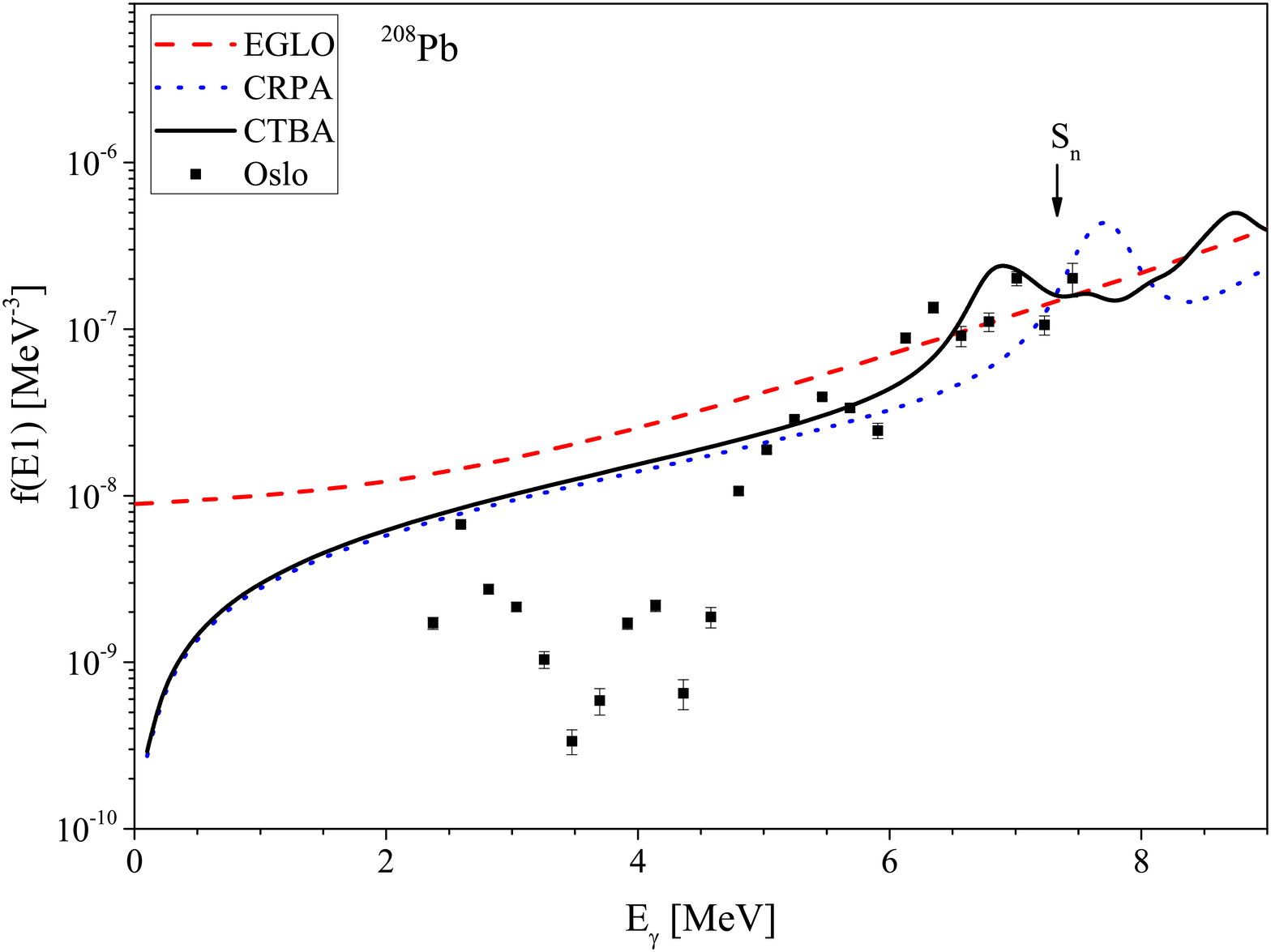}
\caption{Color online. E1 PSF for $^{208}$Pb. Dotted lines correspond to the self-consistent CRPA,  solid lines to the CTBA  (including PC),  and dashed lines to the EGLO model \cite{RIPL2}. Experimental data are taken from Ref. \cite{Oslo}.}
\label{fig-2}       
\end{figure}

In Fig.\ref{fig-1} and Fig.\ref{fig-2} we show the PSFs for $^{132}$Sn and $^{208}$Pb  calculated within  our microscopic (ETFFS(TBA)),  or simply TBA,  and RPA methods with Skyrme forces SLy4. These PSFs have been  recalculated from the theoretical photoabsorption cross sections taken  from  \cite{ave2011}   
  ($^{132}$Sn) and  \cite{TselLut} ($^{208}$Pb). The phenomenological EGLO PSFs are also shown.
In Fig.\ref{fig-1},  the E1 PSF for $^{132}$Sn is compared with experimental data from Ref. \cite{Adrich}. In Fig.\ref{fig-2},  the E1 PSF for $^{208}$Pb is compared with the 
experimental data obtained within the Oslo method \cite{Oslo}. As it can be seen in contrast to the phenomenological model EGLO and microscopic CRPA, the  CTBA approach, i.e. RPA + phonon coupling, 
can describe some observed structures of PDR in $^{132}$Sn and partly $^{208}$Pb only due to  phonon coupling. For $^{132}$Sn,  we see the well-known structure at about 10 MeV (our approach gives a lower energy), usually called as pygmy-dipole resonance for the photoabsorption cross section,  see the discussions in \cite{ave2011, savran2014,  paar2007}.

Let us discuss the results shown in Fig.\ref{fig-2} for $^{208}$Pb. We see that the CTBA approach describes the experiment on the whole at E>5 MeV and  does  it better than CRPA  (note that the smoothing parameter 200 keV has been used in the calculations). 
However,  we have a large  disagreement with experimental data at E<5 MeV.  As one can see  from \cite{rezaeva}, where the transitions between ground end excited states have been measured,  the beginning of the  $1^-$ excitation spectrum in \cite{rezaeva} is  4.84 MeV,  i.e. there is no $1^-$ transitions between ground and excited states below 4.84 MeV. 
This result is  understandable: in the double-magic $^{208}$Pb there is no single-particle  or two-phonon E1 transitions at about E<5 MeV. 

\begin{figure}
\centering
\sidecaption
\includegraphics[width=8cm, clip]{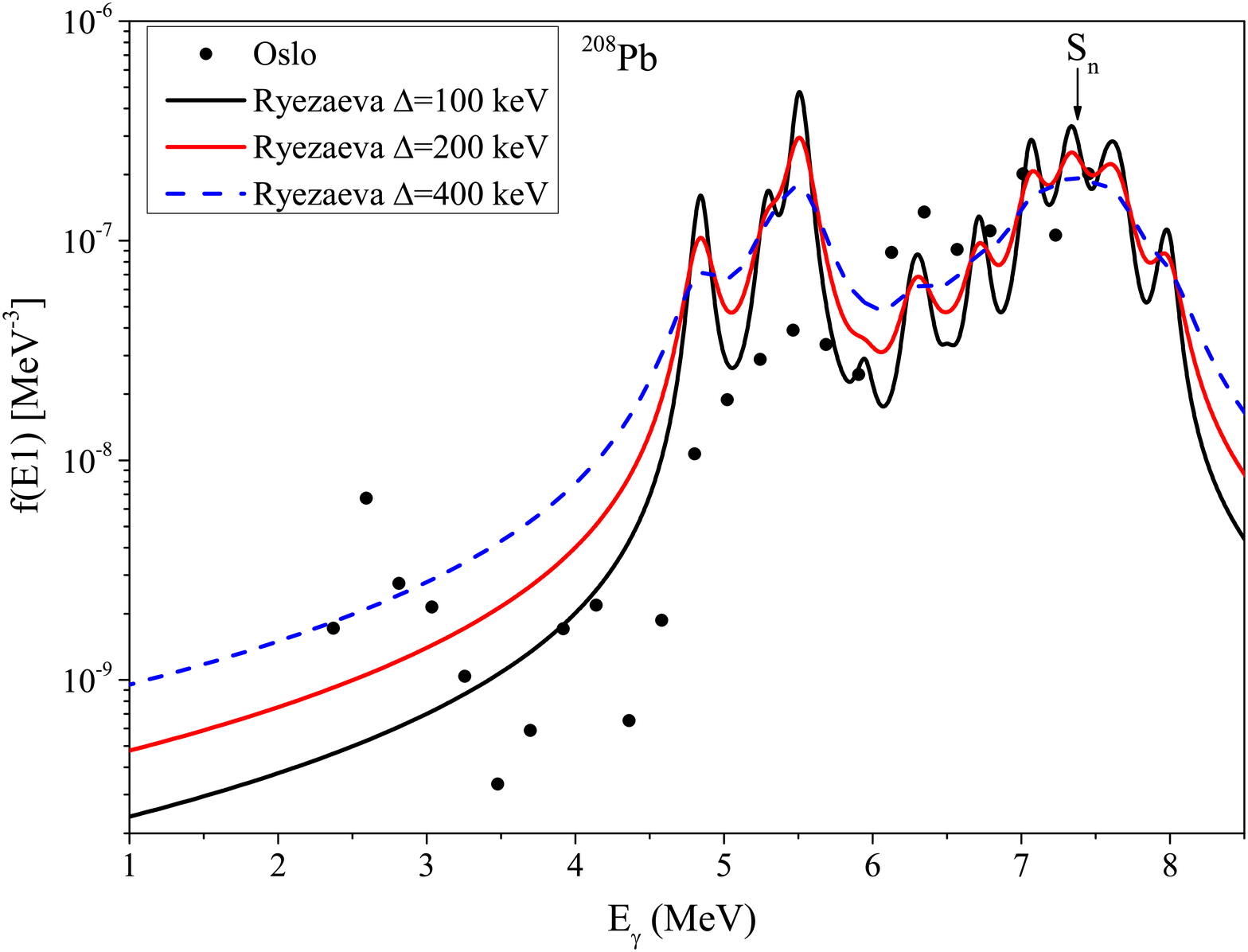}
\caption{Color online. Comparison  of the experimetal  data: the $(^{3}He,^{3}He\acute{}\gamma)$ reactions method \cite{Oslo} and nuclear resonance fluorescence technique \cite{rezaeva}. The lowest 1$^-$-level of  the data \cite{rezaeva} is 4.84 MeV. It was  smoothed by us with three smoothing parameters $\Delta$. See text for details.}
\label{fig-3}       
\end{figure}

In order to obtain some  additional information  we  have compared with  each other two sets of experimental data for $^{208}$Pb  (see Fig.\ref{fig-3}) : 1) the PSF data from \cite{Oslo}  where the transitions between  ground and excited states as well as between excited states were measured and 
2) the data \cite{rezaeva} for the B (E1) values for the transitions between only ground and excited states.
  It is necessary to compare  both sets of data  with  approximately the same smoothing. So,  taking into account that the experimental resolution in the experimental data \cite{Oslo}  is about 200 keV,  we smoothed the data  \cite{rezaeva} with three smoothing parameters 100 keV,  200 keV and 400 keV. 
  As it can be expected,  we have obtained a rough agreement  between both sets of experimental data  at 
  $E_{\gamma} > 4.84$ MeV. 
  Thus,  one can assume that the excitations observed in Ref.~\cite{Oslo} at  $E_{\gamma} < 4.84$~MeV are  caused  mainly by transitions between excited states. However,  it is necessary to note that 
  the mechanisms of the reactions used in Refs.~\cite{Oslo} and \cite{rezaeva} are very different and, what is important here,
 the data from the Oslo $^{208}$Pb  experiment \cite{Oslo} may suffer from a factor of 2 uncertainties due to low level density below the particle separation threshold\footnote{ Private communications with the Oslo group}.

\section{Neutron capture cross sections}
\begin{figure}
\centering
\sidecaption
\includegraphics[width=8cm, clip]{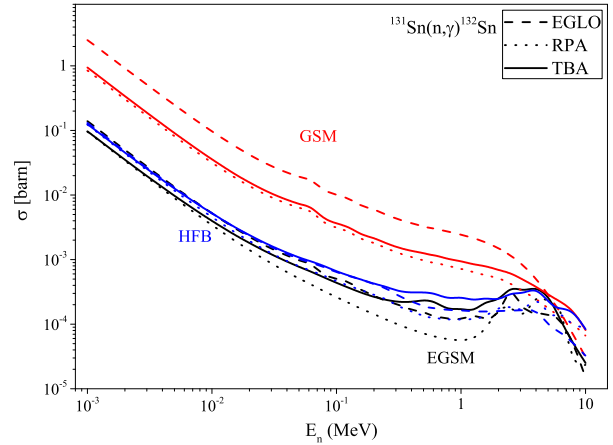}
\caption{Color online.$^{131}Sn (n, \gamma)^{132}Sn$ cross section calculated with the EGLO  (dash),  RPA  (dot) and TBA  (solid) PSFs. The red  curves were calculated using EMPIRE 3.1 with the GSM NLD model,  black ones:  the EMPIRE specific NLD and blue:  the HFB+combinatorial NLD. }
\label{fig-4}       
\end{figure}

\begin{figure}
\centering
\sidecaption
\includegraphics[width=8cm, clip]{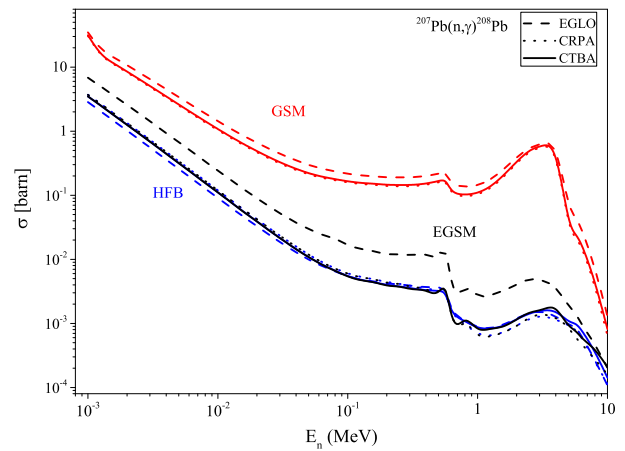}
\caption{Color online.The same as for Fig.\ref{fig-4},  but for the $^{207}Pb (n, \gamma)^{208}Pb$ cross section calculated with the EGLO  (dash),  CRPA  (dot) and CTBA  (solid) PSFs.}  
\label{fig-5}       
\end{figure}

In Fig.\ref{fig-4} and Fig.\ref{fig-5} the neutron radiative capture cross sections are shown for the compound $^{132}$Sn and $^{208}$Pb. Our approach for PSF is non-statistical,  so  there is no sense to compare its results with the available $^{207}Pb (n, \gamma)^{208}Pb$ cross sections
\cite{Green, Wass} because these data  (two points) are  in the neutron resonance energy region.   We see a very large difference between the results obtained with traditional GSM and other NLD models  (EMPIRE specific  and  HFB+combinatorial),  namely,   the difference for $ (n, \gamma)$ cross sections  is about one order of magnitude practically in all the neutron energy up to 2 MeV and 10 MeV for the compound $^{132}$Sn and $^{208}$Pb,  respectively. 
There is no noticeable difference between the results with phenomenological EMPIRE specific and microscopic HFB+combinatorial NLD models. 
One of the possible reasons is that in both cases  the known experimental energies 
of the first $2^+$ levels have been used which is, generally speaking,  important for the phonon enhancement effect in the NLDs  
(it is also useful  to note  that the  energy of the first $2^+$ level in $^{132}$Sn  calculated self-consistently in our TBA is 4.34 MeV while the experimental energy is 4.041 MeV).
A detailed discussion about these results will be presented  somewhere else,  so  we showed so many curves on each of the Fig.\ref{fig-4} and Fig.\ref{fig-5} in order to obtain general information. 
 
\section{Average radiative widths}

\begin{table*}
\centering
\caption{Average radiative widths $\Gamma_\gamma$  (meV) for s-wave neutrons. Systematics is taken from Ref. \cite{Mughab}.}
\label{tab-1}
\renewcommand{\tabcolsep}{0.1cm}
\begin{tabular}{lllllll}
\hline
\multirow{2}{*}{Nuclei}&\multirow{2}{*}{NLD model}&\multirow{2}{*}{EGLO}&\multirow{2}{*}{RPA}&\multirow{2}{*}{TBA}&\multirow{2}{*}{System.}&M1\\
                       &                          &                     &&&                        &contrib.\\\hline
\multirow{3}{*}{$^{132}$Sn}& GSM & 398 & 133 & 148 &   & 40.9 \\
 &EMPIRE specific &7340&4675&5186&&515.3 \\
 & comb. HFB& 4444 & 4279 & 4259 & &340.7 \\
\hline
\multirow{3}{*}{$^{208}$Pb}& GSM & 10.56 & 4.44 & 4.61 &5070  & 0.79 \\
 & EMPIRE specific &6292&2562&2109&3770 &6.56 \\
 & comb. HFB&2734&2973&2448&&5.25 \\
\hline
\end{tabular}
\end{table*}

Unfortunately,  the experimental data are very scarce for double-magic nuclei $^{132}$Sn and $^{208}$Pb. 
However,  for $^{208}$Pb with EMPIRE 3.1 we found, see Table \ref{tab-1}, for the average radiative widths $\Gamma_{\gamma}$ values, a reasonable  agreement  with the systematics \cite{Mughab} only for EMPIRE specific and  HFB + combinatorial NLD models.  
For the average resonance s-wave level spacings $D_0$,  the following  was found: $D_0 (GSM) 
 = 0.00441 \: keV$ 
 ,  $D_0 (EMPIRE \: specific)
 = 32.0 \: keV$,  $D_0 (HFB)
 = 37.6 \: keV$, 
  while $D_0 (exp) = 30 \: (8)\: keV$. The EMPIRE produces unreasonably small value for $D_0 (GSM)$.  

In the last column of Table \ref{tab-1},  the  contribution of M1 resonance \cite{RIPL2} to $\Gamma_{\gamma}$ calculated with EMPIRE 3.1  is given,  which  is based on the standard Lorentz approximation with the width $\Gamma$ = 4 MeV. It  turned out rather small. As discussed in \cite{kaev2006}, this $\Gamma$ value  is very  questionable, especially for $^{208}$Pb.

\section{Conclusion}
Here,  the self-consistent microscopic approach for the PSFs calculations  has been used for the double-magic nuclei $^{132}$Sn and $^{208}$Pb. To calculate neutron radiative cross sections and average radiative widths,  we have used the EMPIRE 3.1 code.
A noticeable specificity of the considered double-magic nuclei has been found. 
The contribution of the phonon coupling is not so noticeable,  on the whole,  as compared with the  semi-magic nuclei \cite{ave2011, Acha2015}.
For  the considered characteristics,   a very significant disagreement between the results obtained with the phenomenological  GSM and microscopic  HFB  NLD models has been found.  The discrepancies between the results with the phenomenological EGLO PSF  and microscopic RPA (or CRPA) or TBA (or CTBA) are much less for the same NLD model.  

The results obtained confirm the necessity of using  consistent microscopic approaches for calculations of radiative nuclear characteristics in double-magic nuclei.
 Also,   due to comparison  of the two sets of experimental data
 \cite{Oslo} and \cite{rezaeva}, it was possible to conclude that the nature of the PSF values  observed in \cite{rezaeva} at E<4.84 MeV for $^{208}$Pb should  be only caused  by the transitions between excited states. 
 
 The authors acknowledge Drs. A. Voinov and  V.G. Pronyaev for useful discussions. V.T. acknowledges Saint Petersburg State University  for the research grant 11.38.648.2013. The authors acknowledge discussions of the $^{208}$Pb results with Therese Renstrom and Profs. S. Siem, M. Guttormsen and A.C. Larsen from the Oslo group.


%
%
%

\end{document}